\documentclass[aps,prl,twocolumn,a4paper,showpacs,superscriptaddress,longbibliography]{revtex4-1}

\usepackage{graphicx}		
\usepackage{dcolumn}		
\usepackage{amsmath,amssymb}		
\usepackage{bm}			
\usepackage{changes}		

\begin{document}

\newcommand{\mfs}{Mn$_{1-x}$Fe$_{x}$Si}
\newcommand{\mcs}{Mn$_{1-x}$Co$_{x}$Si}
\newcommand{\fcs}{Fe$_{1-x}$Co$_{x}$Si}
\newcommand{\cso}{Cu$_{2}$OSeO$_{3}$}


\title{Response of the skyrmion lattice in MnSi to cubic magnetocrystalline anisotropies}

\author{T.~Adams}
\affiliation{Physik Department, Technische Universit\"at M\"unchen, D-85748 Garching, Germany}

\author{M.~Garst}
\affiliation{Institute for Theoretical Physics, Universit\"at zu K\"oln, D-50937 K{\"o}ln, Germany}
\affiliation{Institut f\"ur Theoretische Physik, Technische Universit\"at Dresden, 01062 Dresden, Germany}

\author{A.~Bauer}
\affiliation{Physik Department, Technische Universit\"at M\"unchen, D-85748 Garching, Germany}

\author{R.~Georgii}
\affiliation{Physik Department, Technische Universit\"at M\"unchen, D-85748 Garching, Germany}
\affiliation{Heinz Maier-Leibnitz Zentrum (MLZ), Technische Universit\"at M\"unchen, D-85747 Garching, Germany}

\author{C.~Pfleiderer}
\affiliation{Physik Department, Technische Universit\"at M\"unchen, D-85748 Garching, Germany}

\date{\today}

\begin{abstract}
We report high-precision small angle neutron scattering of the orientation of the skyrmion lattice in a spherical sample of MnSi under systematic changes of the magnetic field direction. For all field directions the skyrmion lattice may be accurately described as a triple-$\vec{Q}$ state, where the modulus $\vert \vec{Q} \vert$ is constant and the wave vectors enclose rigid angles of $120^{\circ}$. Along a great circle across $\langle 100\rangle$, $\langle 110\rangle$, and $\langle 111\rangle$ the normal to the skyrmion-lattice plane varies systematically by $\pm3^{\circ}$ with respect to the field direction, while the in-plane alignment displays a reorientation by $15^{\circ}$ for magnetic field along $\langle 100\rangle$. Our observations are qualitatively and quantitatively in excellent agreement with an effective potential, that  is determined by the symmetries of the tetrahedral point group $T$ and includes contributions up to sixth-order in spin-orbit coupling, providing a full account of the effect of cubic magnetocrystalline anisotropies on the skyrmion lattice in MnSi.
\end{abstract}


%

\maketitle

Magnetocrystalline anisotropies (MCA) are of central importance for the character and detailed properties of magnetic order on all scales. Well-known examples include the easy and hard axis of long-range magnetic order, spin-flop transitions in antiferromagnets, or the formation of magnetic domains. Yet, despite of their inherent importance, the determination of MCAs on a purely experimental level has been very challenging for many decades, in particular for complex magnets with a spatially modulated magnetization.

A particularly intriguing showcase of MCAs was predicted nearly three decades ago in chiral magnets, when Bogdanov and Yablonskii suggested that skyrmion lattices, as derived in terms of solitonic (particle-like) spin configurations, could be stabilized in an applied magnetic field under moderate easy-plane anisotropies, i.e., terms that are second order in spin-orbit coupling (SOC) \cite{1989:Bogdanov:SovPhysJETP,1994:Bogdanov:JMagnMagnMater}. It was, in turn, a major surprise, when skyrmion lattices were first identified experimentally in the B20 compounds MnSi \cite{2009:Muhlbauer:Science}, {\fcs} \cite{2010:Munzer:PhysRevB}, FeGe \cite{2010:Yu:Nature} and related systems \cite{2010:Pfleiderer:JPhysCondensMatter,2012:Seki:Science,2012:Adams:PhysRevLett,2015:Tokunaga:NatCommun}, featuring cubic MCAs that are fourth and sixth order in SOC and therefore much weaker. Moreover, it was shown \cite{2009:Muhlbauer:Science,2013:Buhrandt:PhysRevB,Bauer_book_16} that, instead of the MCAs, stabilization in these systems arises from entropic effects due to thermal fluctuations, where the skyrmion lattice is well represented by a multi-$\vec Q$ magnetic texture \cite{2011:Adams:PhysRevLett}.

While these seminal studies suggested that cubic MCAs play a sub-leading role for the global understanding of skyrmion lattices in cubic chiral magnets, several break-through discoveries have recently identified a number of new directions of research. Examples include the formation of topological defects, such as disclinations, as a new avenue for spintronics applications \cite{2017:Bauer:PRB, 2018:Schoenherr:NatPhys}, and the first observation of two independent skyrmion phases in the same material \cite{2018:Chacon:NatPhys}. Moreover, cubic MCAs are also held responsible for the anisotropy of the temperature and field range of skyrmion phases \cite{Lamago:2006gd,2009:Muhlbauer:Science,2011:Adams:PhysRevLett,2013:Bauer:PhysRevLett}, deviations from an ideal hexagonal lattice under uniaxial strain \cite{Chacon2015,2015:Nii:NatCommun,ShibataK.2015}, the formation of different skyrmion lattice morphologies \cite{2016:Karube:NatMater,2017:Nakajima}, as well as the formation, size, and shape of skyrmion lattice domains \cite{2009:Muhlbauer:Science, 2010:Munzer:PhysRevB,2010:Adams:JPhysConfSer, 2012:Adams:PhysRevLett, Seki:2012ch,2016:Zhang:NanoLett, 2017:Makino:PRB, 2017:Bannenberg:PRB}. Further, cubic MCAs are also essential for the motion of skyrmion lattices under spin currents \cite{2010:Jonietz:Science,Everschor:2012je,Mochizuki:2014fj}. Last but not least, MCAs provide an important point of reference for the understanding of superconducting vortex lattices, where different morphologies \cite{2006:Laver:PRL,2009:Muehlbauer:PRL} and the associated topological character \cite{Laver:2010cl} attract great current interest, while higher-order contributions in the superconducting order parameter of the underlying Ginzburg-Landau theories are still not known.

From a technical point of view accurate determination of MCAs in modulated structures is difficult as the spins point in various crystallographic directions, reflecting a balance of varying contributions. Thus, the effect of MCAs on modulated structures cannot simply be determined by means of torque magnetometry representing the standard technique used in ferromagnets. Moreover, the MCA may not only affect the propagation direction of the modulations but also specific details of the modulations causing changes of pitch and harmonicity. In turn, a given material might even appear to change its easy axis under applied magnetic fields \cite{2018:Chacon:NatPhys,2018:Halder:PRB}. In comparison, imaging methods such as magnetic force \cite{2013:Milde:Science,2016:Milde:NanoLett} or scanning x-ray microscopy \cite{2016:Zhang:NanoLett,2016:Zhang:PhysRevB}, while capable of detecting modulated structures and the associated textures, are sensitive to surface effects. Taken together, this underscores the need for controlled microscopic measurements of MCAs in bulk samples which minimize the influence of the sample shape.

In this Letter we report the determination of the cubic MCAs in MnSi up to sixth order in SOC by means of high-precision small angle neutron scattering (SANS) \cite{Adams:thesis,2017:Bauer:PRB}. Using a spherical sample we minimized parasitic effects of the sample shape. Covering systematically a large range of field directions, we determined the equilibrium orientation of the skyrmion lattice in excellent qualitative and quantitative agreement with an effective potential as inferred from the associated Ginzburg-Landau theory and the symmetries of the tetrahedral point group $T$ \cite{SOM}. Remarkably, for all field directions the skyrmion lattice may be accurately described as a triple-$\vec{Q}$ state, where the modulus $\vert \vec{Q} \vert$ is constant and the wave vectors enclose rigid angles of $120^{\circ}$, i.e., the MCAs have a minor influence on the modulated state as such and mainly affect its orientation. This provides an important benchmark of the response of skyrmion lattices under cubic MCAs. On a more general note, our study demonstrates that systematic neutron diffraction experiments are ideally suited to track the effects of MCAs in materials featuring modulated magnetic structures. 


For our study a single crystal sample was prepared from an optically float-zoned ingot \cite{2011:Neubauer:RevSciInstrum}. The single crystal was carefully ground into a sphere with a diameter of 5.75\,mm. The sample shape was chosen to minimize inhomogenities of the demagnetization fields, known from previous work to cause parasitic deviations of the skyrmion lattice orientation \cite{2009:Muhlbauer:Science,2011:Adams:PhysRevLett,2018:Reimann:PRB}. The sample was oriented by means of x-ray Laue diffraction to better than a few degrees and attached to an aluminum sample holder. The precise orientation of the sample during the SANS measurements was inferred from the position of the intensity maxima in the helical state, which from numerous independent studies are known to be accurately aligned along the $\langle111\rangle$ directions at zero field \cite{1976:Ishikawa:SSC, 1977:Ishikawa:PRB, 2010:Janoschek:PRB, 2017:Bauer:PRB}. 

The SANS measurements were carried out at the diffractometer MIRA~\cite{2015:Georgii:JLSFR,2018:Georgii:NIMA} at the Heinz Maier-Leibnitz Zentrum~(MLZ) using a neutron wavelength $\lambda = (10.4 \pm 0.5)\,\mathrm{\AA}$. The distance between sample and detector was 2.1\,m. Two apertures with a distance of 1.5\,m and an opening of $2 \times 2\,\mathrm{mm}^{2}$ were placed between the monochromator and the sample. The azimuthal and radial resolution of our set up were $\Delta \alpha_{\mathrm{az}} = 3.8^{\circ}$ and $\Delta q = 0.0024\,\mathrm{\AA}^{-1}$, respectively. A closed-cycle cryostat equipped with a rotatable sample stick was used to cool the sample, while permitting rotation of the sample with respect to the vertical axis. The applied magnetic field was generated with a pair of Helmholtz coils. 


\begin{figure}
\includegraphics[width=1.0\linewidth]{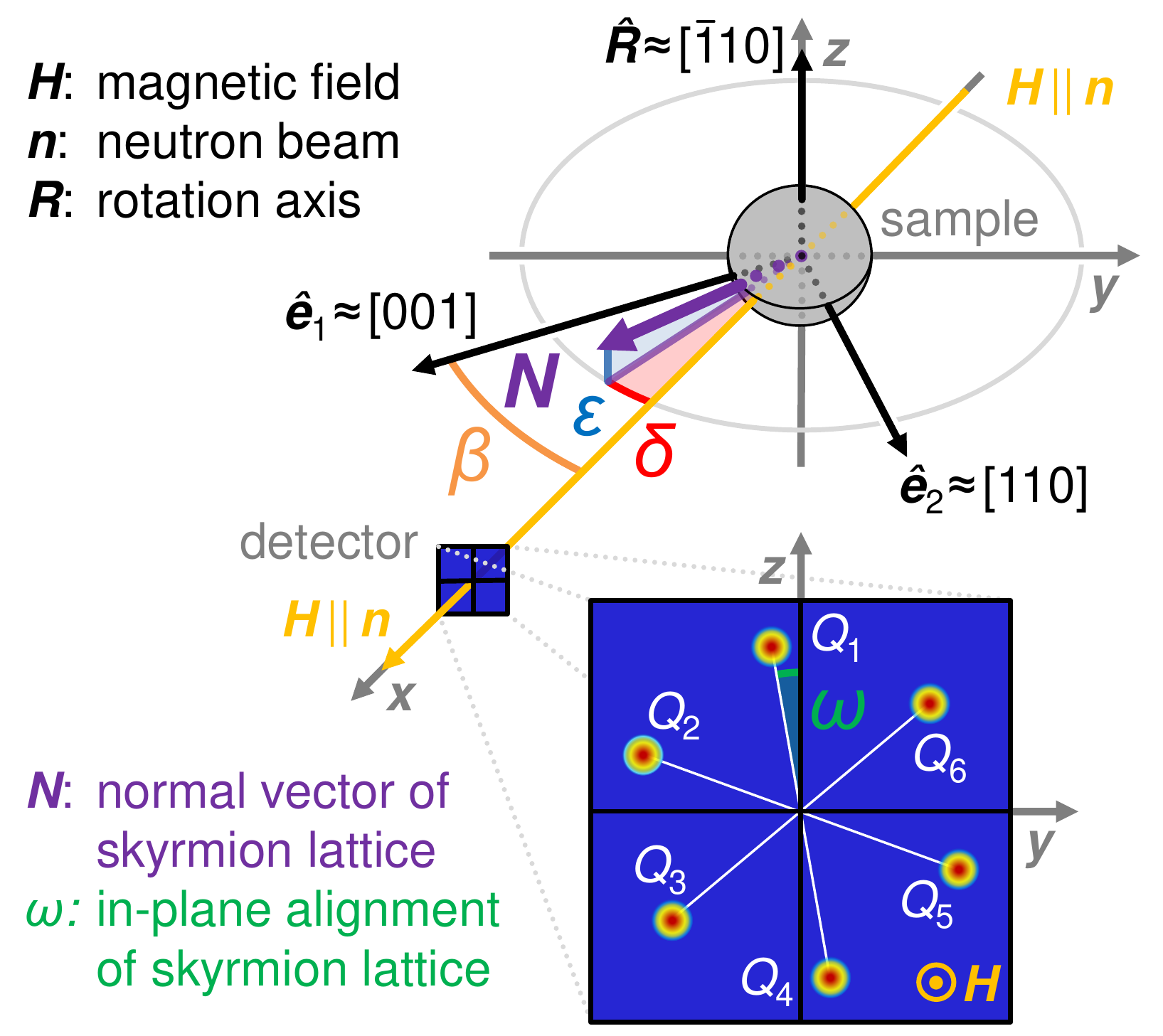}
\caption{Schematic illustration of the neutron scattering setup. 
The magnetic field $\bm{H}$ was applied parallel to the neutron beam $\bm{n}$. The sample could be rotated with respect to the axis $\hat R$, where the rotation angle is denoted by $\beta$. Deviations of the normal vector $\bm{N}$ of the skyrmion lattice from $\bm{H}$ are described by the angles $\varepsilon$ and $\delta$, accounting for the tilt towards $\hat R$ and $\hat H \times \hat R$, respectively.
The orientation of the sixfold scattering pattern within the plane perpendicular to $\bm{N}$ is described by $\omega$, denoting the angle between the rotation axis $\hat R$ and the closest wavevector $\bm{Q}_1$ of the Bragg pattern. In our experiments, the rotation axis $\hat R$ pointed along $[\bar{1} 1 0]$ up to a small misalignment described in the text, thus $\hat e_1 \approx [001]$ and $\hat e_2 \approx [110]$.
}
\label{figure1}
\end{figure}

Shown in Fig.~\ref{figure1} is a schematic illustration of the angles required to describe our experimental set-up and data. The sample could be rotated with respect to the vertical axis described by $\hat R$. A small misalignment of $\hat R$ with respect to the $[\bar{1} 10]$ axis of the sample is described by the angles $\theta_r$ and $\phi_r$, such that $\hat R = (-\cos \theta_r \cos (\phi_r - \pi/4), -\cos \theta_r \sin(\phi_r - \pi/4), \sin \theta_r)$. For $\theta_r = \phi_r = 0$ the rotation axis corresponded to the crystallographic $[\bar{1} 10]$ direction; finite angles $\theta_r$ and $\phi_r$ describe tilts of the rotation axis towards $[001]$ and $[\bar{1}\bar{1}0]$, respectively. 
The direction of the magnetic field, $\hat H = \hat e_1 \cos \beta + \hat e_2 \sin \beta$ was perpendicular to $\hat R$, where $\hat e_1 = (\cos (\phi_r - \pi/4) \sin \theta_r, \sin (\phi_r - \pi/4) \sin \theta_r, \cos \theta_r)$ and $\hat e_2 = \hat R \times \hat e_1$. Rotations of the sample with respect to $\hat R$ are described by the angle $\beta$, such that $\hat H = (\sin \beta/\sqrt{2},\sin \beta/\sqrt{2},\cos\beta)$ for $\theta_r = \phi_r = 0$. 

The orientation of the sixfold diffraction pattern of the skyrmion lattice involves three degrees of freedom, namely the angles $\delta$, $\varepsilon$, and $\omega$. The first two describe the orientation of the skyrmion lattice plane as defined by the unit normal vector 
$\hat N = \hat H \cos\varepsilon \cos\delta+ \hat R \sin \varepsilon + (\hat H \times \hat R)\cos \varepsilon \sin \delta$.
$\hat N$ is approximately parallel to the magnetic field, $\hat H$. Small tilts towards $\hat R$ and $\hat H \times \hat R$ are described by $\varepsilon$ and $\delta$, respectively. Ignoring these small tilts, the orientation of the diffraction peaks are described by six unit vectors $\hat Q_n = \hat R \cos(\omega + (n-1) \pi/3) + (\hat H \times \hat R) \sin(\omega + (n-1)\pi/3)$ within the plane, where $\omega$ is the angle  between $\hat Q_1$ and the direction of the rotation axis $\hat R$. 

In our study, we determined the detailed orientation of the skyrmion lattice for a large number of applied magnetic field orientations, $\beta$, covering half of a great circle in steps of $5^{\circ}$. For each angle $\beta$ the sample was cooled in an applied magnetic field of 183\,mT from well above $T_{c}$ to the center of the skyrmion lattice phase pocket at 28.3\,K. Subsequently, we performed a horizontal ($\pm3.5^{\circ}$) as well as a vertical ($\pm2^{\circ}$) rocking scan, both with a step size of $0.25^{\circ}$ with respect to the direction of the incoming neutron beam (see \cite{SOM} for technical details on the uneven intensity distribution). At the end of each measurement the sample was heated up and $\beta$ was changed to the next value. 

\begin{figure}
\includegraphics[width=1.0\linewidth]{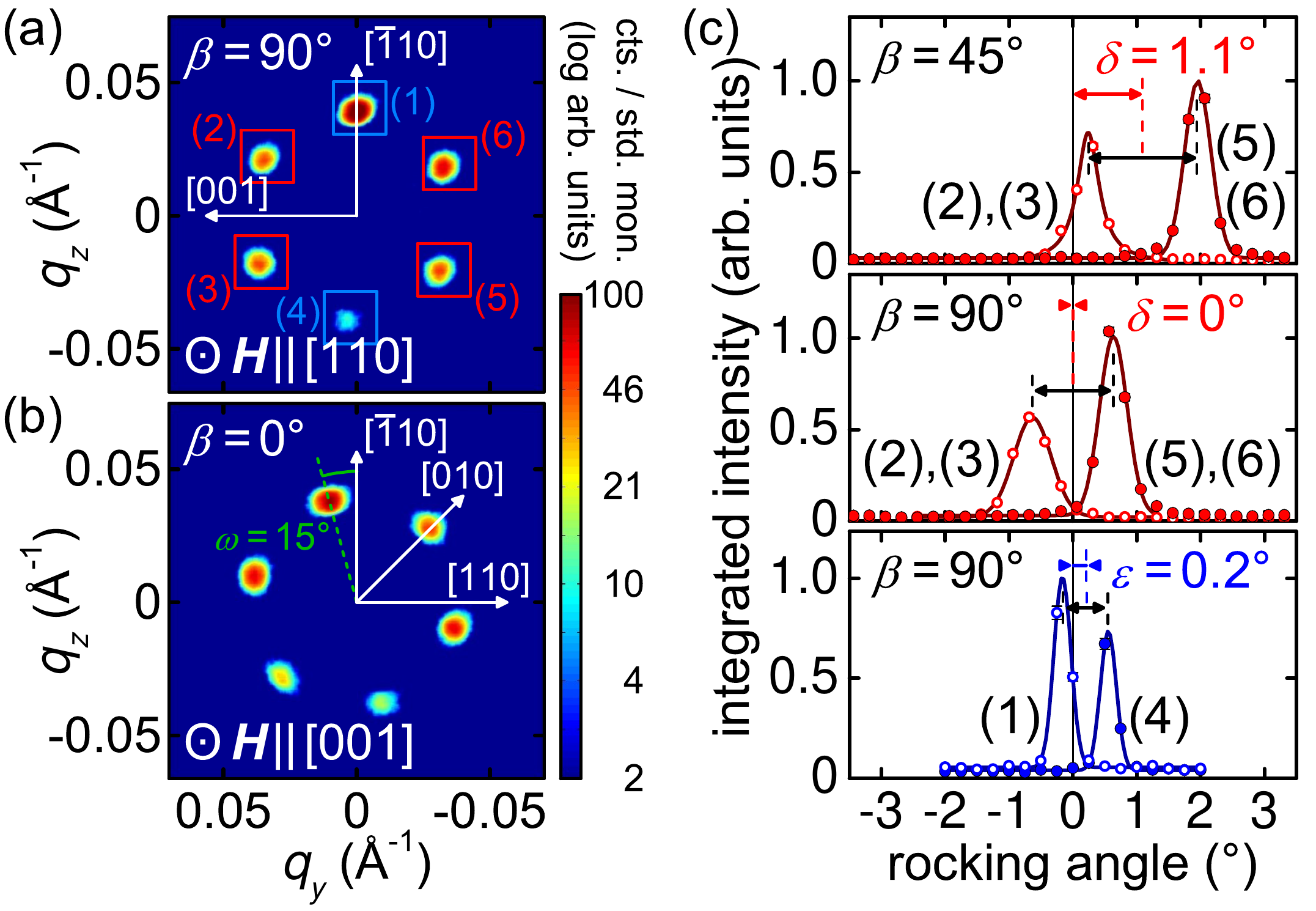}
\caption{Typical small-angle neutron scattering data and analysis of these data. \mbox{(a),(b)}~Sum over horizontal and vertical rocking scans for magnetic field along $[110]$ ($\beta = 90^{\circ}$) and $[001]$ ($\beta = 0^{\circ}$) showing the sixfold scattering pattern of the skyrmion lattice. For technical reasons the intensity distribution is uneven; see \cite{SOM} for details. (c)~Integrated intensity of the red and blue boxes in the SANS data [cf. panel (a)] as a function of the vertical and horizontal rocking angle, respectively. Solid lines represent Gaussian fits. The angles $\delta$ and $\varepsilon$ correspond to the arithmetic means of the peak centers.}
\label{figure2}
\end{figure}

For all field directions studied the sum over both rocking scans showed the characteristic sixfold scattering pattern of the skyrmion lattice state, as depicted in Figs.~\ref{figure2}(a) and \ref{figure2}(b) \cite{SOM}. However, the orientation of the pattern in reciprocal space displayed subtle changes. To track these changes we labeled the intensity maxima of $Q_{1}$ through $Q_{6}$ counter-clockwise, starting at the top with the maximum closest to the rotation axis, $\hat R$. In order to determine the normal vector of the skyrmion lattice, $\hat N$, we first considered the integrated intensity of the maxima $Q_{2}$, $Q_{3}$, $Q_{5}$, and $Q_{6}$ (red boxes) as a function of the vertical rocking angle, see top and middle panel of Fig.~\ref{figure2}(c) for $\beta = 45^{\circ}$ and $\beta = 90^{\circ}$. Data were well-described by Gaussian profiles where the arithmetic mean of the rocking centers provided the azimuthal deviation angle $\delta$. Analogously, the integrated intensity of the maxima $Q_{1}$ and $Q_{4}$ (blue boxes) as a function of the horizontal rocking angle provided the polar deviation $\varepsilon$ as shown for $\beta = 90^{\circ}$ in the bottom panel of Fig.~\ref{figure2}(c). The in-plane alignment of the sixfold scattering pattern was finally described by the angle $\omega$ between the rotation axis and the closest intensity maximum. It was extracted from Gaussian fits to the integrated intensity of the sum over both rocking scans as a function of the azimuthal angle (not shown). 

\begin{figure}
\includegraphics[width=1.0\linewidth]{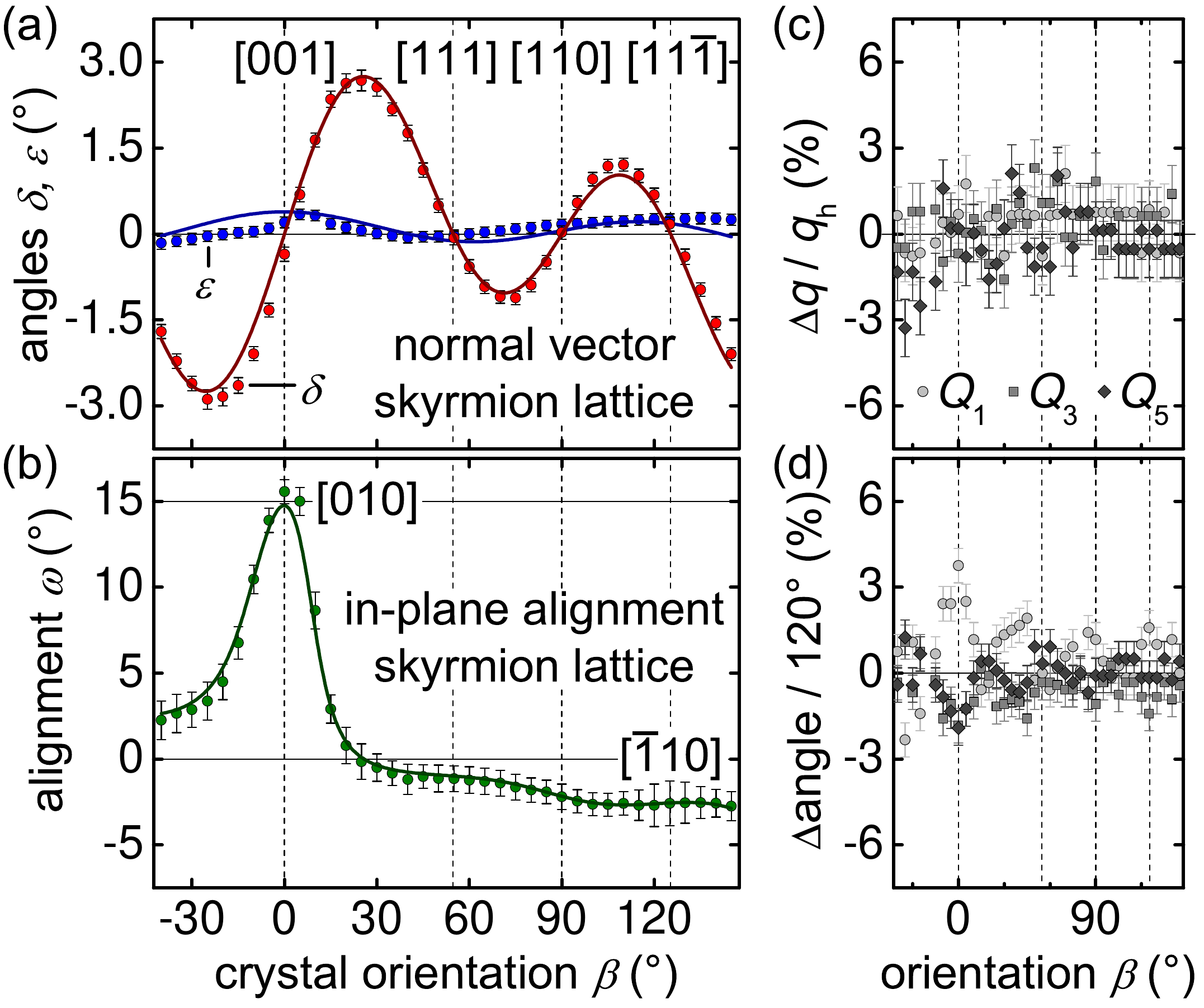}
\caption{Orientation of the skyrmion lattice as a function of the field direction as parametrized by $\beta$. Specific directions of $\bm{H}$ are indicated at the top of panel (a) ignoring the tiny misalignment of $\bm{R}$. (a)~Deviation of the normal vector $\hat N$ of the skyrmion lattice from the magnetic field orientation as parametrized by $\delta$ and $\varepsilon$. (b)~Alignment of the sixfold scattering pattern in the skyrmion lattice plane specified by $\omega$. For $\hat R = [\bar{1} 1 0]$ at $\omega = 0^\circ$ and $\omega = 15^\circ$ one of the Bragg wavevectors is aligned along $[010]$ and $[\bar{1}10]$, respectively. Solid lines are fits that account for the small misalignment of the rotation axis from $[\bar{1} 1 0]$, see text.\mbox{(c),(d)}~Variation of the modulus of the Bragg vectors and their enclosed angles. Within the error bars the Bragg pattern remains rigidly sixfold symmetric.
}
\label{figure3}
\end{figure}

Our experimental data on the orientation of the skyrmion lattice as a function of the field direction are summarized in Fig.~\ref{figure3}. The error bars shown in Fig.~\ref{figure3} were determined as part of the fitting procedure described above and reflect the high accuracy at which changes of the orientation of the skyrmion lattice could be determined. Panel (a) shows the orientation of the skyrmion lattice plane. The deviation $\delta$ of its normal from the field direction within the rotation plane vanishes for fields along major crystallographic axes. These configurations were covered in previous experiments \cite{2009:Muhlbauer:Science,2011:Adams:PhysRevLett,2017:Bannenberg:PRB}, unless badly aligned samples were studied. For fields away from the major orientations, the anisotropies caused a systematic meandering with $\delta$ reaching values in the range $\pm3^{\circ}$. In comparison, the deviation with respect to the rotation axis, $\varepsilon$, was less than $0.4^{\circ}$. 

The in-plane orientation $\omega$ as a function of field direction is shown in Fig.~\ref{figure3}(b). For most directions we observe only a small deviation $\omega$, implying that $\bm{Q}_{1}$ is approximately aligned with $[\bar{1}10]$. This is consistent with previous reports \cite{2009:Muhlbauer:Science, 2010:Jonietz:Science, 2011:Adams:PhysRevLett} where fields were aligned along $\langle110\rangle$. However, for a magnetic field close to $[001]$, i.e., $\beta=0^{\circ}$, a pronounced reorientation takes place such that $\omega \approx 15^{\circ}$. Here one of the $\bm{Q}$ vectors of the Bragg pattern is aligned along $[010]$, see also Fig.~\ref{figure2}(b). As depicted in Figs.~\ref{figure3}(c) and \ref{figure3}(d), neither the modulus of the wave vectors $Q_{1}$, $Q_{3}$, and $Q_{5}$ nor the angle they enclosed changed as a function of the applied field direction within experimental accuracy. This implies that the skyrmion lattice for any orientation of the magnetic field is characterized by a rigid sixfold Bragg pattern.


The observation of a rigid skyrmion lattice allows to reduce the theoretical interpretation from the full Ginzburg-Landau theory to a minimal model of a rigid skyrmion lattice in a potential that takes into account the symmetries of the skyrmion lattice and the tetrahedral point group $T$ \cite{2009:Muhlbauer:Science,2010:Munzer:PhysRevB,2017:Bauer:PRB}. For information on the relationship of the Ginzburg-Landau parameters with the coefficients of this potential we refer to the supplementary information \cite{SOM}.

We first discuss the orientation of the unit normal vector $\hat N$ as described by the potential
\begin{equation} \label{PotentialNormal}
\mathcal{V}(\hat N) = - \frac{k}{2}(\hat N \hat H)^2 + \mu (\hat N_x^4 + \hat N_y^4 + \hat N_z^4).
\end{equation}
The first term in $k$ favors an alignment of the normal with the magnetic field, i.e., it represents the tilting energy in the absence of MCAs, whereas the second term in $\mu$ accounts for changes of the tilting energy due to the MCAs. In the limit of small $\mu/k \ll 1$, one expects only a slight deviation of the normal from the magnetic field orientation. The contribution of the crystalline potential in $\mu$ yields a small force that slightly tilts $\hat N$ away from $\hat H$. One obtains $\hat N = \hat H$ when this force vanishes which is the case for directions where the potential displays either an extremum or a saddle point, i.e., for field along $\langle 111\rangle$, $\langle 001\rangle$ and $\langle 110\rangle$. Perturbatively minimizing the potential in the limit of small $\mu$, $\theta_r$, and $\phi_r$ one obtains for the deviations in lowest order

\begin{align}
\delta &\approx \frac{\mu}{2 k} [1 + 3 \cos(2 \beta)] \sin(2 \beta),\\
\varepsilon &\approx 
\frac{\mu}{2 k}\Big(8 \phi_r \sin^3 \beta + \theta_r [3 \cos \beta + 5 \cos(3 \beta)]\Big).
\end{align}
The polar deviation $\varepsilon$ vanishes exactly in case of perfect alignment of the rotation axis $\hat R$ with $[\bar{1}10]$, i.e., $\theta_r = \phi_r = 0$.

This brings us to the orientation of the skyrmion lattice within the plane orthogonal to $\hat N$. Due to its sixfold symmetry, the orientation of the skyrmion lattice is governed by terms that are sixth order in the wavevectors $\bm{Q}_{j}$ and, therefore, sixth-order in spin-orbit coupling \cite{2010:Munzer:PhysRevB}. There are two independent sixth order invariants for a unit vector $\hat{Q}$ that are consistent with the tetrahedral point group $T$, notably: 
$I_{1}(\hat{Q}) = \hat{Q}_x^6 + \hat{Q}_y^6 + \hat{Q}_z^6$ and $I_{2}(\hat{Q}) = \hat{Q}_x^2 \hat{Q}_y^4 + \hat{Q}_y^2 \hat{Q}_z^4 + \hat{Q}_z^2 \hat{Q}_x^4$. An effective potential for the skyrmion lattice orientation $\omega$ is obtained by summing over the invariants of the sixfold symmetric Bragg wavevectors, i.e., $\mathcal{V}(\omega) = \sum_{n=1}^6 \sum_{i=1}^2 \lambda_i I_{i}(\hat{Q}_n)$. Its explicit form is given by 
\begin{equation} \label{omegaPot}
\mathcal{V}(\omega) = A_{\hat R}(\beta) \cos (6\omega) + B_{\hat R}(\beta) \sin (6\omega),
\end{equation}
where the coefficients $A$ and $B$ depend on the rotation axis $\hat R$ and on the rotation angle $\beta$. 

For a perfect alignment of $\hat R$ with $[\bar 1 10]$, i.e., for $\theta_r = \phi_r = 0$ where $\hat H = (\sin \beta/\sqrt{2},\sin \beta/\sqrt{2},\cos\beta)$ the coefficients simplify to

\begin{align}
A_{[\bar 1 10]} &= -  \frac{9 \lambda_1}{1024} [5 \sin \beta + \sin(3\beta)]^2,\\
B_{[\bar 1 10]} &= -  \frac{3 \lambda_2}{1024} [66 \cos \beta - 11 \cos(3\beta) + 9 \cos(5\beta)].
\end{align}
For a magnetic field along $[110]$, i.e., $\beta = \pi/2$, the coefficient $B$ vanishes and the potential is solely determined by the $\cos (6\omega)$ dependence. The $\bm{Q}_1$ vector of the Bragg pattern is then predicted to be aligned along $[\bar 1 10]$ with $\omega = 0$ for positive $\lambda_1 > 0$, but $\omega$ is shifted by $30^\circ$ for negative $\lambda_1 < 0$ so that one of the $\bm{Q}$ vectors is instead aligned along $[001]$. 
A sign change of $\lambda_1$ as a function of temperature/field and pressure thus accounts for the transition observed in Cu$_2$OSeO$_3$ \cite{Seki:2012ch} and MnSi \cite{Chacon2015}, respectively. For a magnetic field along $[001]$, i.e., $\beta = 0$, the $A$ coefficient vanishes and the potential is proportional to $\sin(6\omega)$. In this case, one of the $\bm{Q}$ vectors aligns along $[010]$ or $[100]$ for positive $\lambda_2 > 0$ or negative $\lambda_2 < 0$, respectively. For MnSi at ambient pressure we find $\lambda_1,\lambda_2 > 0$.

After accounting for the various signs, the result for $\omega$ inferred from a minimization of the potential depends on the ratio $\lambda_2/\lambda_1$. For MnSi we obtain for the angle $\omega(\beta) = \frac{1}{6} \arctan(B_{[\bar 1 10]}/A_{[\bar 1 10]}) + \delta \omega$ where the perturbative correction $\delta \omega$ is due to the small misalignment of the rotation axis. To lowest order in $\theta_r$ and $\phi_r$ it is given by  

\begin{align}
\delta \omega =& \frac{A_{[\bar 1 10]} \delta B - B_{[\bar 1 10]} \delta A}{6 (A^2_{[\bar 1 10]}+B^2_{[\bar 1 10]})},\\
\delta A =& - \frac{9 \lambda_2}{2048} \Big[\theta_r (143 \sin(2\beta) - 44 \sin(4\beta) + 3 \sin(6\beta)) 
\nonumber\\
&+
\phi_r (154+77 \cos(2\beta) + 22 \cos(4\beta) + 3 \cos(6\beta)) 
\Big],
\\
\delta B  = &\frac{9 \lambda_1}{256} (5\sin \beta + \sin(3\beta))
\nonumber\\&\times
\Big[\theta_r (5-\cos(2\beta)) + 4 \phi_r \sin(2\beta) 
\Big].
\end{align}

In order to compare theory and experiment, we included $\phi_r$ and $\theta_r$ as fitting parameters since the experiment is particularly sensitive to a small misalignment of the rotation axis. In outstanding agreement with experiment the theory accounts for the  orientation of the skyrmion lattice observed as described by $\delta$, $\varepsilon$, and $\omega$, as shown by the solid lines in Figs.~\ref{figure3}(a) and \ref{figure3}(b). From a combined fit we obtain for the misalignment angles $\phi_r = 0.4^\circ$ and $\theta_r = 2.2^\circ$ as well as for the parameters of the effective potentials $\mu/k = 0.04$ and $\lambda_2/\lambda_1 = 0.14$.

In summary, our SANS measurements provide a detailed qualitative and quantitative account of the precise skyrmion lattice orientation in agreement with MCAs up to sixth order in spin-orbit coupling in accordance with the tetrahedral symmetry group $T$ of MnSi. Interestingly, the potential Eq.\,\eqref{omegaPot} for the in-plane orientation $\omega$ vanishes for twelve directions of the magnetic field which, for the particular value $\lambda_2/\lambda_1 = 0.14$, are specified by $\hat H \approx \pm (0.98,0.21,0)$ and symmetry-equivalent directions. On the unit sphere spanned by $\hat H$, these points constitute singularities with winding number $w = 1/6$. 

We predict that encircling each of these points on the unit sphere as a function of field orientation, will generate a rotation of the Bragg pattern by an angle $\Delta \omega  = 2\pi w = \pi/3$. This will provide a critical test of the so-called ''hairy ball theorem", which has been discussed in the analogous context of the Abrikosov vortex lattice in type II superconductors \cite{Laver:2010cl}. A related open question to be addressed experimentally in the future concerns whether these singular points account for the degenerate multidomain configurations of the skyrmion lattice that have been observed in some of the cubic chiral magnets \cite{2010:Munzer:PhysRevB,2016:Milde:NanoLett,2016:Zhang:NanoLett,2017:Bannenberg:PRB}.


We wish to thank A.~Rosch, P.~B\"{o}ni, S. M{\"u}hlbauer and M. Laver for helpful discussions. Financial support through DFG TRR80 (From Electronic Correlations to Functionality; projects E1 and F2), DFG FOR960 (project P4), DFG SPP2137 (Skyrmionics), ERC AdG 291079 (TOPFIT), and ERC AdG 788031 (ExQuiSid) is gratefully acknowledged. M.G. was supported by  DFG SFB1143 (Correlated Magnetism: From Frustration To Topology) and DFG grant 1072/5-1.


%

\end{document}


\newcommand{\todo}[1]{\textbf{\textsc{\textcolor{red}{(TODO: #1)}}}}
\newcommand{\OLD}[1]{{\tiny {\bf old:} #1 }}
\newcommand{\NEW}[1]{{ \it #1 }}
\renewcommand{\vec}[1]{\boldsymbol{#1}}
\newcommand{\w}{\omega}

\newcommand{\fcs}{Fe$_{1-x}$Co$_{x}$Si}
\newcommand{\mfs}{Mn$_{1-x}$Fe$_{x}$Si}
\newcommand{\mcs}{Mn$_{1-x}$Co$_{x}$Si}
\newcommand{\cso}{Cu$_{2}$OSeO$_{3}$}

\newcommand{\rxx}{$\rho_{\rm xx}$}
\newcommand{\rxy}{$\rho_{\rm xy}$}
\newcommand{\rxyt}{$\rho_{\rm xy}^{\rm T}$}
\newcommand{\Drxyt}{$\Delta\rho_{\rm xy}^{\rm top}$}
\newcommand{\Sxy}{$\sigma_{\rm xy}$}
\newcommand{\Sxya}{$\sigma_{\rm xy}^A$}

\newcommand{\bco}{$B_{\rm c1}$}
\newcommand{\bct}{$B_{\rm c2}$}
\newcommand{\bao}{$B_{\rm A1}$}
\newcommand{\bat}{$B_{\rm A2}$}
\newcommand{\beff}{$B^{\rm eff}$}

\newcommand{\btr}{$B^{\rm tr}$}

\newcommand{\tc}{$T_{\rm c}$}
\newcommand{\ttr}{$T_{\rm tr}$}

\newcommand{\mb}{$\mu_0\,M/B$}
\newcommand{\dmdb}{$\mu_0\,\mathrm{d}M/\mathrm{d}B$}
\newcommand{\ddmddb}{$\mathrm{\mu_0\Delta}M/\mathrm{\Delta}B$}
\newcommand{\cm}{$\chi_{\rm M}$}
\newcommand{\cac}{$\chi_{\rm ac}$}
\newcommand{\rechi}{${\rm Re}\,\chi_{\rm ac}$}
\newcommand{\imchi}{${\rm Im}\,\chi_{\rm ac}$}

\newcommand{\ozz}{$\langle100\rangle$}
\newcommand{\ooz}{$\langle110\rangle$}
\newcommand{\ooo}{$\langle111\rangle$}
\newcommand{\too}{$\langle211\rangle$}

\makeatletter 
\renewcommand{\thefigure}{S\@arabic\c@figure}
\makeatother
\setcounter{secnumdepth}{2}

\title{Supplementary information for:\\ 
Response of the skyrmion lattice in MnSi to cubic magnetocrystalline anisotropies}

\author{T.~Adams}
\affiliation{Physik Department, Technische Universit\"at M\"unchen, D-85748 Garching, Germany}

\author{M.~Garst}
\affiliation{Institute for Theoretical Physics, Universit\"at zu K\"oln, D-50937 K{\"o}ln, Germany}
\affiliation{Institut f\"ur Theoretische Physik, Technische Universit\"at Dresden, 01062 Dresden, Germany}

\author{A.~Bauer}
\affiliation{Physik Department, Technische Universit\"at M\"unchen, D-85748 Garching, Germany}

\author{R.~Georgii}
\affiliation{Physik Department, Technische Universit\"at M\"unchen, D-85748 Garching, Germany}
\affiliation{Heinz Maier-Leibnitz Zentrum (MLZ), Technische Universit\"at M\"unchen, D-85747 Garching, Germany}

\author{C.~Pfleiderer}
\affiliation{Physik Department, Technische Universit\"at M\"unchen, D-85748 Garching, Germany}

\date{\today}

\begin{abstract}
We report supplementary information on the relationship between the anisotropy potential used in the main text for the analysis of our experimental data and the Ginzburg-Landau theory of cubic chiral magnets. We also provide further information on the experimental methods, notably typical rocking scans and intensity patterns observed in small angle neutron scattering.
\end{abstract}


\vskip2pc

\maketitle


\section{Derivation of the anisotropy potential}

The Ginzburg-Landau theory for cubic chiral magnets represents an effective low-energy theory in terms of the magnetization $\vec{M}$. Detailed introductions may, for instance, be found in Ref.\,\onlinecite{1980:Bak:JPhysCSolidState,1980:Nakanishi:SolidStateCommun,2009:Muhlbauer:Science,2010:Munzer:PhysRevB}. It is important to note that the small spin-orbit coupling $\lambda_{\rm SOC}$ observed in MnSi implies a hierarchy of interactions. Up to second order in $\lambda_{\rm SOC}$, the Ginzburg-Landau theory includes the ferromagnetic exchange, the Dzyaloshinskii-Moriya interaction, as well as the dipolar interaction. On this level it is invariant with respect to a combined $SO(3)$ rotation in real and spin space. In the presence of an applied magnetic field, $\vec H$, this symmetry is reduced to a $O(2)$ rotation symmetry around the axis of the magnetic field $\vec H$. Taking additionally into account terms that are fourth order and higher order in  spin-orbit coupling, the rotation symmetry is reduced by magnetocrystalline anisotropies (MACs). These are of central interest for the study reported in the main text.

An analysis of the impact of MCAs on the skyrmion lattice in terms of the full Ginzburg-Landau theory for the magnetization is a challenging task. In order to capture quantitatively the response of the modulated magnetization with respect to changes in the magnetic field, the differential susceptibility tensor needs to be evaluated which is generally not accessible on the mean-field level. However, it turns out that the description simplifies considerably in the limit of small spin-orbit coupling, which is applicable to MnSi. 

To lowest order in spin-orbit coupling MCAs may be neglected, and the skyrmion lattice ground state possesses a perfect six-fold symmetry characterized by a normal unit vector $\hat N$ describing the orientation of the skyrmion lattice plane and an angle $\omega$ that accounts for the orientation of the skyrmion lattice within this plane. In equilibrium, the normal $\hat N$ is aligned with the magnetic field direction. A tilt of $\hat N$ away from this direction will cost energy and we demonstrate below that this energy cost is quantified by a single quantity, i.e., the stiffness $k$ that may be expressed in terms of the homogeneous magnetization and the transverse susceptibility evaluated at fixed $\hat N$. To lowest-order in spin-orbit coupling the angle $\omega$ remains undetermined due to the rotation symmetry around the applied magnetic field. 

The effect of MCAs may then be captured by contributions to the effective potentials for $\hat N$ and $\omega$ that are governed by the tetrahedral point group symmetry. The coefficients $\mu$, $\lambda_1$ and $\lambda_2$ of the resulting potentials may be related to corresponding coefficients of MCAs in the Ginzburg-Landau theory by perturbatively evaluating them within the approximate mean-field solution in the absence of MCAs, which possesses the six-fold symmetry. This approach is borne out of our experimental observation of the modulus and angular relationship of the ${\bf Q}$-vectors, indicating that the distortion of the skyrmion lattice for different field directions remains negligible. The response of the skyrmion lattice is finally addressed by minimizing the effective potentials for $\hat N$ and $\omega$ as discussed in the mains text. 

As a remark on the side, the same general strategy has also been used to account for the reorientation transition between the helix and conical phases in cubic chiral magnets, which may be addressed in terms of an effective potential for the orientation of the single-${\bf Q}$ helical modulation \cite{2017:Bauer:PRB}.

The presentation is organized as follows. We begin in section \ref{tilt-1} with a discussion of the tilting stiffness of the skyrmion lattice, i.e., changes of orientation of $\hat N$, in the absence of MCAs as expressed by the coefficient $k$. This is followed in section \ref{tilt-2} by a discussion of the tilting stiffness in the presence of MCAs, motivating the additional parameter $\mu$. Finally, the in-plane orientation will be addressed in section \ref{rotate}, providing insights on the parameters $\lambda_1$ and $\lambda_2$ .


\subsection{Skyrmion lattice tilting stiffness without MCAs}
\label{tilt-1}

We begin with the free energy of a skyrmion lattice with a fixed normal vector $\hat N$ in the \textit{absence of magnetocrystalline anisotropies}. For symmetry reasons the associated energy density is given by
\begin{align} \label{FreeEnergy}
F(\hat N) = - \frac{k}{2} (\hat N \hat H)^2 + F_0,
\end{align}
where $\hat H = \vec H/H$ denotes the direction of the magnetic field. The coefficient $k$ introduced here and in the context of the potential given in the main text are identical. It quantifies the tilting stiffness of the skyrmion lattice with respect to changes of direction of $\hat N$. In turn, the coefficient $k$ as well as the contribution $F_0$ to the free energy depend on the strength of the magnetic field $H = |\vec H|$, i.e., $k = k(H)$ and $F_0 = F_0(H)$. In equilibrium and in the absence of MCAs the free energy $F(\hat N)$ of the skyrmion lattice is minimal for $\hat N = \hat H$.

The coefficient $k$ may be related to the differential magnetic susceptibility of the skyrmion lattice phase distinguishing the longitudinal and transverse susceptibilities, $\chi_{\parallel}$ and $\chi_{\perp}$, respectively. The longitudinal susceptibility is given by $\chi_{\parallel}=\partial M/\partial H$, where $M = |\vec M|$ is the magnitude of the magnetization. In contrast, for the transverse susceptibility, describing changes of the magnetization under a perpendicular perturbing field, two limits must be considered. Namely, the susceptibility for unconstrained changes of the direction of the normal vector $\hat N$, denoted $\chi_{\perp, \rm free}$, and the susceptibility when the direction of the normal vector $\hat N$ is fixed and cannot change, denoted $\chi_{\perp, \rm fixed}$. In comparison to the transverse susceptibility, the longitudinal susceptibility is the same for both unconstrained and fixed normal vector $\hat N$.

If the normal vector is unconstrained and free to change its orientation, it adopts its optimal value for $\hat N = \hat H$ instantaneously under changes of field direction, and the magnetization $\vec M$ is always parallel to the field, $\vec M = M \hat H$. For this unconstrained response the susceptibility tensor is given by the standard expression
\begin{align}
\frac{\partial \vec M_i}{\partial \vec H_j}\Big|_{\hat N\, {\rm free}} = 
\frac{\partial M}{\partial H} \hat H_j \hat H_i + \frac{M}{H} (\delta_{i j} - \hat H_i \hat H_j),
\end{align}
and the associated unconstrained transverse susceptibility becomes 
\begin{equation}
\chi_{\perp, \rm free} = \frac{M}{H} 
\end{equation}
However, as shown in detail below, for constrained (fixed) normal vector $\hat N$ the susceptibility tensor is given by
\begin{align} \label{SuscFixedN}
\frac{\partial \vec M_i}{\partial \vec H_j}\Big|_{\hat N\, {\rm fixed}, \hat N \to \hat H} = 
\frac{\partial M}{\partial H} \hat H_j \hat H_i + \chi_\perp (\delta_{i j} - \hat H_i \hat H_j),
\end{align}
where the limit $\hat N \to \hat H$ was taken after taking the derivative and the transverse susceptibility is given by 
\begin{align} \label{SuscPerp}
\chi_{\perp, \rm fixed} = \frac{M}{H} - \frac{k}{\mu_0 H^2}.
\end{align}
Thus, the tilting stiffness $k$ in the absence of MCAs may be inferred from the difference of the transverse susceptibilities for unconstrained and constrained (fixed) normal vector $\hat N$, notably 
\begin{align}
k = \mu_0 H^2 \left(\frac{M}{H} - \chi_{\perp, \rm fixed} \right).
\end{align}

It is now interesting to note that the transverse susceptibility $\chi_{\perp, \rm fixed}$ may, in principle, be determined experimentally by means of an ac susceptibility measurement, where a small oscillating field component perpendicular to the applied static magnetic field stabilizing the skyrmion lattice generates an oscillating change of orientation $\vec H(t)$. In practice the direction of the oscillating field component has to be chosen such that the effects of MCAs vanish. The response of the skyrmion lattice as described by the normal vector $\hat N(t)$ then reflects to what extent the skyrmion lattice is capable of following any changes of the field direction as a function of time. 

In fact, the relaxation of the normal vector $\hat N$ into its equilibrium position will require a relaxation time $\tau$ that may be expected to be large as it involves the reorientation of macroscopic skyrmion lattice domains. Thus, for relatively large ac frequencies $\omega \tau \gg 1$, the normal vector $\hat N$ remains essentially unaffected. In contrast, $\hat N(t)$ will be able to follow the changes of field orientation for low frequencies $\omega \tau \ll 1$. It is therefore expected that the transverse ac susceptibility assumes the values $\chi_{\perp, \rm fixed}$ and $\chi_{\perp, \rm free}=M/H$ in the limits of very large and very low frequencies, respectively. These consideration compare with the frequency dependence of the ac susceptibility at the helix reorientation transition, which is also governed by very long relaxation times \cite{2017:Bauer:PRB}. Thus, the relevant behavior for the skyrmion lattice may be accessible experimentally.


Returning now to the expression for $\chi_{\perp, \rm fixed}$ given in Eq.~\eqref{SuscPerp}, we present in the following a derivation starting with the magnetization associated with the free energy density $F$ in Eq.~\eqref{FreeEnergy} at a fixed $\hat N$,
\begin{equation}
\begin{split}
& \mu_0 \vec M_i(\hat N) = - \frac{\partial F}{\partial \vec H_i} \\
& =  \frac{k'}{2} (\hat N \hat H)^2 \hat H_i + k (\hat N \hat H) \frac{\hat N_j}{H} (\delta_{ji} - \hat H_j \hat H_i) - F_0' \hat H_i
\\&
= \Big[(\hat N \hat H)^2 \left(\frac{k'}{2}  - \frac{k}{H}\right) - F_0' \Big] \hat H_i + k \frac{\hat N \hat H}{H} \hat N_i ,
\end{split}
\end{equation}
where $k' = \partial_H k$ and $F'_0 = \partial_H F_0$. In equilibrium $\hat N = \hat H$, and the magnetization is parallel to the applied magnetic field 
\begin{align} \label{Magnetization}
\mu_0 \vec M_i(\hat N = \hat H) =  \left(\frac{k'}{2} -  F_0' \right) \hat H_i .
\end{align}
Computing the differential susceptibility at fixed $\hat N$ we find
\begin{align}
\begin{split}
\mu_0 \frac{\partial \vec M_i(\hat N)}{\partial \vec H_j} = &
\mu_0 \frac{\partial \vec M_i(\hat N)}{\partial H} \hat H_j \\
&+ \mu_0 \frac{\partial \vec M_i(\hat N)}{\partial \hat H_\ell}   \frac{1}{H}(\delta_{\ell j} - \hat H_\ell \hat H_j).
\end{split}
\end{align}
This expression comprises two parts that originate in the dependences on $H$ and $\hat H$, i.e., the strength and the orientation of the magnetic field, respectively. Considering these contributions separately, we obtain
\begin{equation}
\begin{split}
 \mu_0 \frac{\partial \vec M_i(\hat N)}{\partial H} 
 & =
 \left((\hat N \hat H)^2 \left(\frac{k''}{2}  - \frac{k'}{H} + \frac{k}{H^2}\right) - F_0'' \right) \hat H_i \\
& + (\hat N \hat H) \hat N_i \Big(\frac{k'}{H} - \frac{k}{H^2}\Big), 
\end{split}
\end{equation}
and
\begin{align}
\begin{split}
\mu_0 \frac{\partial \vec M_i(\hat N)}{\partial \hat H_\ell} 
&=  
\left((\hat N \hat H)^2 \left(\frac{k'}{2}  - \frac{k}{H}\right) - F_0' \right) \delta_{i\ell} \\
&+ 2 (\hat N \hat H) \left(\frac{k'}{2}  - \frac{k}{H}\right)   \hat H_i  \hat N_\ell
+ \frac{k}{H} \hat N_\ell \hat N_i .
\end{split}
\end{align}
In the limit $\hat N \to \hat H$, the susceptibility tensor simplifies to
\begin{equation}
\begin{split}
&\mu_0 \frac{\partial \vec M_i}{\partial \vec H_j}\Big|_{\hat N\, {\rm fixed}, \hat N \to \hat H} 
= \left(\frac{k''}{2} - F_0''\right) \hat H_j \hat H_i \\
&+ \frac{1}{H}\left[\left(\frac{k'}{2}  - \frac{k}{H}\right)  - F_0' \right] (\delta_{i j} - \hat H_i \hat H_j).
\end{split}
\end{equation}
Using the result for the magnetization given in Eq.\,\eqref{Magnetization}, notably $M = \frac{k'}{2} - F'_0$, we obtain the expression for $\chi_{\perp, \rm fixed}$ given in Eq.~\eqref{SuscPerp}.


\subsection{MCA potential of skyrmion lattice tilting}
\label{tilt-2}

Having considered the tilting stiffness of the skyrmion lattice in the absence of MCAs, we turn next to the tilting stiffness in the presence of the MCAs. In the potential $\mathcal{V}$ this is accounted for by the terms with the additional parameter $\mu$. The magnetocrystalline anisotropies entering the Ginzburg-Landau theory of chiral magnets such as MnSi must satisfy the tetrahedral point group symmetries consisting of a threefold $C_3$ rotation symmetry around the crystallographic $\langle 111 \rangle$ axes and a twofold $C_2$ rotation symmetry around crystallographic $\langle 100 \rangle$ axes. Here one has to take into account contributions that break the continuous rotation symmetry both in real and spin space. For the tilting of the skyrmion lattice plane, described by the normal vector $\hat N(t)$, it is instructive to consider the following three terms
\begin{equation} \label{4aniso}
\begin{split}
\mathcal{F}_{4} = 
& K_1 (\vec M_x^4 + \vec  M_y^4 + \vec M_z^4) \\
&+ K_2 \vec M (\partial_x^4 + \partial_y^4 + \partial_z^4 ) \vec M \\
& + K_3 [(\partial_x \vec M_x)^2 +  (\partial_y \vec M_y)^2 + (\partial_z \vec M_z)^2 ].
\end{split}
\end{equation}
Whereas the first term breaks the continuous rotational symmetry in spin space, the second term breaks this symmetry in real space. The third term, finally, breaks this symmetry simultaneously in spin and real space. 

In order to classify these contributions according to powers of spin-orbit coupling $\lambda_{\rm SOC}$, it is important to note that each derivative eventually contributes a power of $\lambda_{\rm SOC}$ as the Dzyaloshinskii-Moriya interaction leads to spatial variations of the magnetization on a length scale $\sim 1/\lambda_{\rm SOC}$. Whereas the coefficients themselves differ in their magnitude $K_1 \sim \mathcal{O}(\lambda^4_{\rm SOC})$, $K_2 \sim \mathcal{O}(\lambda^0_{\rm SOC})$, and $K_3 \sim \mathcal{O}(\lambda^2_{\rm SOC})$, all three terms essentially contribute only at fourth order in spin-orbit coupling due to the derivatives. The situation is thus rather more involved than in conventional ferromagnets without modulations where only the first term would be relevant. In the following, we will refrain from providing an exhaustive list of all terms at a given order in spin-orbit coupling and refer instead to the literature, see for example Ref.~\onlinecite{Karinphdthesis}.


Turning now to the relationship of the Ginzburg-Landau theory with the potential given in the main text, we revisit the effective Landau potential for the normal vector $\hat N$ of the skyrmion lattice at lowest order, 
\begin{equation} \label{Naniso}
\mathcal{V}_{\rm aniso}(\hat N) = \mu (\hat N_x^4 + \hat N_y^4 + \hat N_z^4),
\end{equation}
with the coefficient $\mu \sim \mathcal{O}(\lambda_{\rm SOC}^4)$. As the normal vector $\hat N$ characterizes an entire skyrmion lattice domain, it is spatially constant. The effective magnetocrystalline potential for $\hat N$ is thus analogous to the that of a ferromagnet with a spatially constant magnetization, i.e.,  Eq.~\eqref{Naniso} is the only term at this order of $\lambda_{\rm SOC}$. For the chiral magnets considered in our paper, the coefficient $\mu$ may in principle be related to the corresponding coefficients of the full Ginzburg-Landau theory for the magnetization. 

To lowest order in spin-orbit coupling $\lambda_{\rm SOC}$, this may be achieved by treating the magnetocrystalline anisotropies in lowest-order perturbation theory. This will be illustrated in the following for the three terms listed in Eq.~\eqref{4aniso}. To zeroth order in the MCAs, the equilibrium magnetization of the skyrmion lattice may be well approximated by a triple-$\vec Q$ structure \cite{2009:Muhlbauer:Science},
\begin{equation} \label{SLAnsatz}
\begin{split}
& \vec M(\vec r) 
=  M_0 \hat H \\
&+ M_Q \sum^3_{i=1} \left(\hat e_{1 i} \cos(Q \hat Q_{i} + \phi_i)
+ \hat e_{2 i} \sin(Q \hat Q_{i} + \phi_i)\right),
\end{split}
\end{equation}
where $M_0$ and $M_Q$ are the amplitudes of the mean and the modulated magnetization. The three sets of unit vectors form an orthogonal frame each, $\hat e_{1i} \times \hat e_{2i} = \hat Q_{i}$ for $i=1,2,3$, and, in addition, $\hat Q_{1} + \hat Q_{2} + \hat Q_{3} = 0$. The phases may be chosen arbitrarily provided that $\phi_1 + \phi_2 + \phi_3 = \frac{\pi}{2}$. Plugging this Ansatz into Eq.~\eqref{4aniso} and averaging over a magnetic unit cell, one obtains an effective anisotropy potential \eqref{Naniso} with coefficient $\mu = \mu_1 + \mu_2 + \mu_3$ where
\begin{align}
\mu_1 &= K_1 M_s^4 \left(\frac{9}{64} + \frac{39}{32} m^2 - 
\frac{23}{64} m^4 - \frac{\sqrt{3}}{6} m (1-m^2)^{3/2} \right), \\
\mu_2 &= K_2 M_s^2 Q^4 \frac{3}{8} (1-m^2), \\
\mu_3 &= K_3 M_s^2 Q^2 \frac{3}{16} (1-m^2).
\end{align}
Here we identified $M_s$ with the averaged magnitude of the magnetization, $\sqrt{\langle {\bf M}^2 \rangle} = \sqrt{M_0^2 + 3 M_Q^2}$, and $m = M_0/M_s$. 

All three  terms of Eq.~\eqref{4aniso} contribute to the coefficient $\mu$. In fact, there are even more contributions accounting for magnetocrystalline anisotropies in the Ginzburg-Landau theory that are not listed in Eq.~\eqref{4aniso}. This observation is important in two respects. On the one hand, a description of the experimental behaviour of the normal vector $\hat N$ on the level of the full Ginzburg-Landau theory is not unique as there are different combinations of magnetocrystalline terms that yield the same value of $\mu$, i.e., the knowledge of $\hat N$ is not sufficient to reconstruct all of these magnetocrystalline contributions. On the other hand, it is the beauty of emergent simplicity that the experimental behavior of $\hat N$ is captured by the effective Landau potential of Eq.~\eqref{Naniso} characterized by a single coefficient $\mu$ given the rather involved structure of the parent Ginzburg-Landau theory for the magnetization. 

Taken together, the tilting of the skyrmion lattice plane may be described by the combination of the parameters $k$ and $\mu$, where $k$ measures the tilting stiffness without MCAs, and $\mu$ measures the strength of the anisotropy potential.


\subsection{MCA potential of skyrmion lattice rotation}
\label{rotate}

For the orientation of the skyrmion lattice within the plane orthogonal to $\hat N$, magnetocrystalline anisotropies of higher order than in Eq.~\eqref{4aniso} are required. Namely, due to the sixfold symmetry of the skyrmion lattice, we need to consider magnetocrystalline anisotropies at least of sixth order in spin-orbit coupling, $\mathcal{O}(\lambda_{\rm SOC}^6)$. In order to keep the discussion simple, we consider only two such contributions to the Ginzburg-Landau theory, namely
\begin{equation} \label{6aniso}
\begin{split}
\mathcal{F}_{6} 
= & K_4 \vec M (\partial_x^6 + \partial_y^6 + \partial_z^6 ) \vec M + \\
&K_5 \vec M (\partial_x^2 \partial_y^4 + \partial_y^2 \partial_z^4 + \partial_z^2 \partial_x^4 ) \vec M .
\end{split}
\end{equation}
The coefficient of both terms are of the order $K_{4,5} \sim \mathcal{O}(\lambda_{\rm SOC}^0)$. However, due to the derivatives they only contribute to order $\mathcal{O}(\lambda_{\rm SOC}^6)$ and may therefore be expected to be particularly weak. The first term, $K_4$, possesses a higher symmetry than the tetrahedral point group as it is invariant with respect to fourfold $C_4$ rotations around the $\langle 100 \rangle$ axes. The second term, $K_5$, has a lower symmetry as it  is invariant only with respect to $C_2$ rotations around $\langle 100 \rangle$.

Treating these terms in lowest-order perturbation theory with the Ansatz in Eq.\,\eqref{SLAnsatz}, we obtain a potential that may be expressed in terms of the three unit vectors $\hat Q_i$, with $i=1,2,3$, characterizing the orientation of the six Bragg peaks of the skyrmion lattice phase,
\begin{equation} \label{Qpot}
\begin{split}
\mathcal{V} = &2 \lambda_1 \sum^3_{i = 1} (\hat Q^6_{i,x} + \hat Q^6_{i,y} + \hat Q^6_{i,z}) \\
&+2 \lambda_2 \sum^3_{i = 1} (\hat Q^2_{i,x} \hat Q^4_{i,y} + \hat Q^2_{i,y} \hat Q^4_{i,z} + \hat Q^2_{i,z} \hat Q^4_{i,x}).
\end{split}
\end{equation}
The two terms in the brackets are the two sixth-order invariants that are consistent with the symmetries of the tetrahedral point group. In the main text, we discussed this potential in terms of six vectors $\pm \hat Q_1, \pm \hat Q_2$ and $\pm \hat Q_3$ and the factor of two in Eq.~\eqref{Qpot} ensures that the definition of the coefficients $\lambda_{1,2}$ are consistent with the main text. Perturbatively, we obtain for these coefficients from the two terms in Eq.~\eqref{6aniso},
\begin{align}
\lambda_1 &=  K_4 M_s^2 Q^6 \left(- \frac{1}{6}\right) (1- m^2),
\\
\lambda_2 &= K_5 M_s^2 Q^6 \left(- \frac{1}{6}\right) (1- m^2).
\end{align}
Note that there are further contributions to $\lambda_{1,2}$ from MCAs not listed in Eq.~\eqref{6aniso} similar to the case of $\mu$ in Eq.~\eqref{Naniso}. 

Experimentally it is found that $\lambda_2/\lambda_1 \approx 0.14$, although both terms are nominally of sixth order in $\lambda_{\rm SOC}$. It is tempting to speculate that this small ratio reflects a relatively weak breaking of the fourfold $C_4$ rotation symmetry in MnSi that lowers the point group from octahedral to tetrahedral symmetry.

\begin{figure*}
\includegraphics[clip, trim=0cm 2cm 0cm 0cm, width=1.00\textwidth]{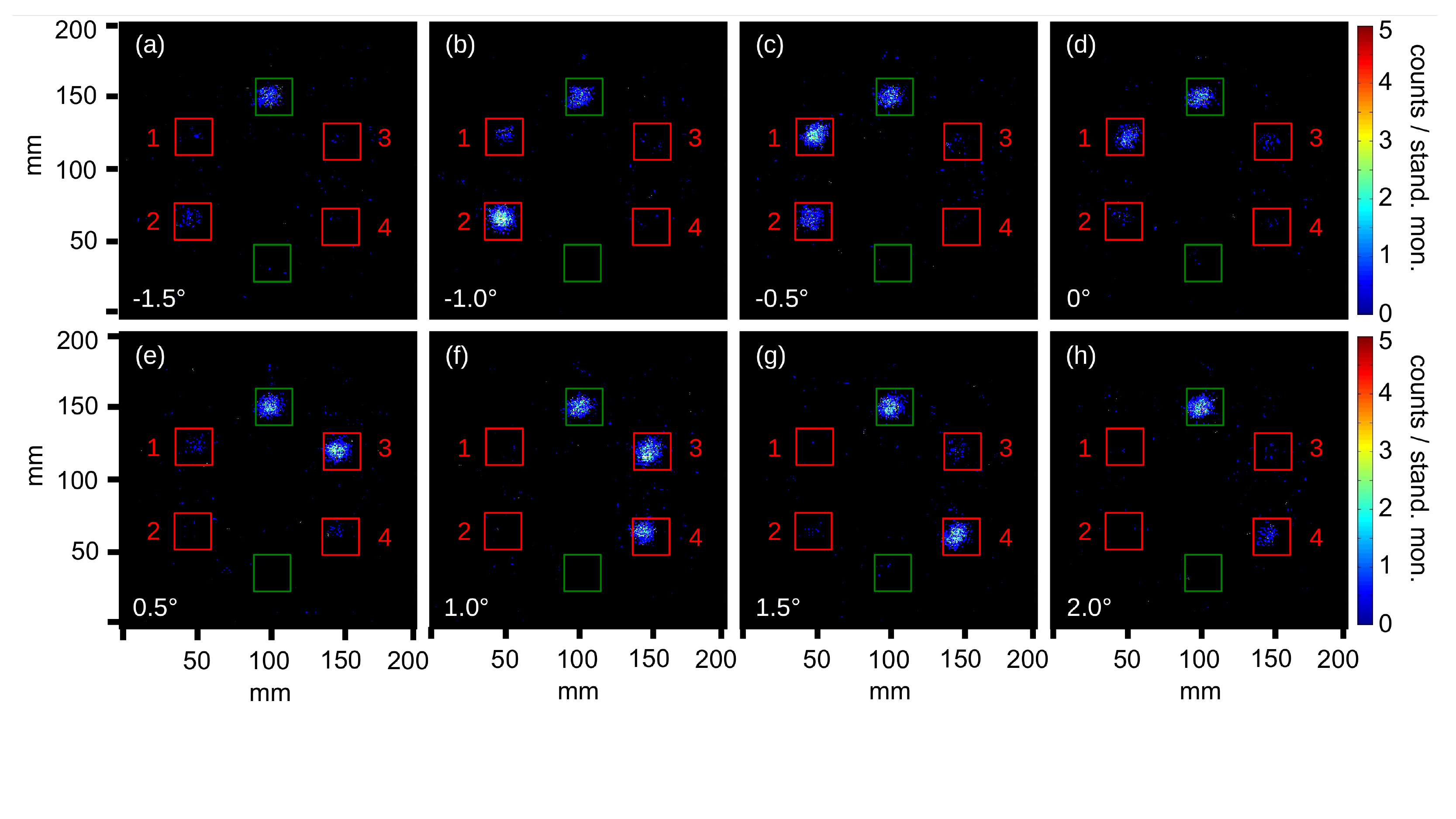}
\caption{Typical scattering patterns as recorded during a horizontal rocking scan in the A-phase of MnSi (the rotation axis is vertical). Data were recorded with a CASCADE detector measuring $200\times200\,{\rm mm^2}$. The rocking scan shown here covered a total of 29 angular positions between -3.5$^{\circ}$ and +3.5$^{\circ}$ in steps of 0.25$^{\circ}$. Due to the very sharp rocking width of the individual peaks, intensity of individual peaks was only observed in a narrow angular range.  Shown here are data for $\beta=90^{\circ}$ and the following rocking angles: (a) -1.5$^{\circ}$, (b) -1.0$^{\circ}$, (c) -0.5$^{\circ}$, (d) 0$^{\circ}$, (e) +0.5$^{\circ}$, (f) +1.0$^{\circ}$, (g) +1.5$^{\circ}$, and (h) +2.0$^{\circ}$.
}
\label{figureS1}
\end{figure*}


\section{Experimental Methods}

For the accurate determination of the orientation of the skyrmion lattice so-called rocking scans were performed with respect to a vertical and a horizontal axis. In a rocking scan the sample and magnetic field are both rotated (rocked) with respect to the position of interest, i.e., the orientation for which the incoming neutron beam and the magnetic field were strictly parallel. This allows to determine the distribution of angles under which the scattering condition is satisfied and provides information on the so-called mosaic spread. 

As our study pursued the collection of a large systematic data set, rocking scans for different field orientations were recorded automatically covering a large number of angular positions. Horizontal rocking scans were performed between -3.5$^{\circ}$ and +3.5$^{\circ}$ in steps of 0.25$^{\circ}$, where typical data are shown in Fig.\,\ref{figureS1}. Vertical rocking scans were performed between -2$^{\circ}$ and +2$^{\circ}$ in steps of 0.25$^{\circ}$. The sum over all horizontal and vertical rocking positions represents the so-called integrated scattering intensity. However, in our study all rocking scans were centered with respect to the field direction. In turn, due to the small tilting of the skyrmion lattice plane, which was only identified during the full analysis of the data after data collection had been completed, not all rocking scans were complete in a strict sense covering all angular positions under which some scattering was still observable. While it was nonetheless possible to determine the direction of $\hat N$ at very high accuracy fitting all peaks simultaneously, the intensity distribution in sums over rocking scans appears to be uneven. It is important to emphasize, that this effect represents a purely technical consequence of the incomplete range of the rocking scans.

For instance, the typical intensity patterns shown in Figs.\,2\,(a) and 2\,(b) in the main text were selected to illustrate the definition and accuracy at which the tilting angles of the skyrmion lattice, $\delta$ and $\epsilon$, and the in-plane rotation angle, $\omega$, could be determined. Unfortunately the tilting angles for these positions was not aligned with the main field direction, hence the intensity distribution as a function of azimuthal angle appears to be slightly uneven.

